\documentclass[draft]{agujournal2018}
\usepackage{xcolor}
\usepackage{natbib}
\usepackage{apacite}
\usepackage{url} %this package should fix any errors with URLs in refs.
\usepackage{lineno}
\journalname{JGR: Planets}

\begin{document}

\title{Early Dynamics of the Lunar Core}

%% ------------------------------------------------------------------------ %%
%
%  AUTHORS AND AFFILIATIONS
%
%% ------------------------------------------------------------------------ %%

% Authors are individuals who have significantly contributed to the
% research and preparation of the article. Group authors are allowed, if
% each author in the group is separately identified in an appendix.)

% List authors by first name or initial followed by last name and
% separated by commas. Use \affil{} to number affiliations, and
% \thanks{} for author notes.
% Additional author notes should be indicated with \thanks{} (for
% example, for current addresses).

% Example: \authors{A. B. Author\affil{1}\thanks{Current address, Antartica}, B. C. Author\affil{2,3}, and D. E.
% Author\affil{3,4}\thanks{Also funded by Monsanto.}}

\authors{Matija {\'C}uk\affil{1}, Douglas P. Hamilton\affil{2}, Sarah T. Stewart\affil{3}}

% \affiliation{1}{First Affiliation}
% \affiliation{2}{Second Affiliation}
% \affiliation{3}{Third Affiliation}
% \affiliation{4}{Fourth Affiliation}

\affiliation{1}{SETI Institute, 189 North Bernardo Avenue, Mountain View, CA 94043}
\affiliation{2}{University of Maryland, College Park}
\affiliation{3}{University of California, Davis}

%(repeat as many times as is necessary)

%% Corresponding Author:
% Corresponding author mailing address and e-mail address:

% (include name and email addresses of the corresponding author.  More
% than one corresponding author is allowed in this LaTeX file and for
% publication; but only one corresponding author is allowed in our
% editorial system.)

% Example: \correspondingauthor{First and Last Name}{email@address.edu}

\correspondingauthor{Matija {\'C}uk}{mcuk@seti.org}

%% Keypoints, final entry on title page.

%  List up to three key points (at least one is required)
%  Key Points summarize the main points and conclusions of the article
%  Each must be 100 characters or less with no special characters or punctuation

% Example:
% \begin{keypoints}
% \item	List up to three key points (at least one is required)
% \item	Key Points summarize the main points and conclusions of the article
% \item	Each must be 100 characters or less with no special characters or punctuation
% \end{keypoints}

\begin{keypoints}
\item The liquid lunar core is dynamically decoupled from the mantle. 
\item In the distant past the mutual obliquity between the lunar core and the mantle was large. 
\item The friction at the core-mantle boundary probaly kept the Moon out of synchronous rotation during periods of high obliquity. 
\end{keypoints}

\begin{abstract}
The Moon is known to have a small liquid core, and it is thought that in the distant past the core may have produced strong magnetic fields recorded in lunar samples. Here we implement a numerical model of lunar orbital and rotational dynamics that includes the effects of a liquid core. In agreement with previous work, we find that the lunar core is dynamically decoupled from the lunar mantle, and that this decoupling happened very early in lunar history. Our model predicts that the lunar core rotates sub-synchronously, and the difference between the core and the mantle rotational rates was significant when the Moon had a high forced obliquity during and after the Cassini State transition. We find that the presence of the lunar liquid core further destabilizes synchronous rotation of the mantle for a wide range of semimajor axes centered around the Cassini State transition. CMB torques make it even more likely that the Moon experienced large-scale inclination damping during the Cassini State transition. We present estimates for the mutual core-mantle obliquity as a function of Earth-Moon distance, and we discuss plausible absolute time-lines for this evolution. We conclude that our results are consistent with the hypothesis of a precession-driven early lunar dynamo and may explain the variability of the inferred orientation of the past lunar dynamo.
\end{abstract}

\section{Introduction}

The Moon has a much smaller bulk density than Earth, indicating that the lunar iron core is much smaller than that of Earth. Lunar laser ranging (LLR) and GRAIL mission results indicate that the Moon's core has a radius of about 340~km (with the 200-380~km diameter range allowed by data) and is (at least partially) liquid \citep{wil06, wil14}. The presence of the core is further suggested by magnetic induction \citep{hoo99, shi13} and seismic modeling \citep{web11, gar11}.  While the Moon does not have a global magnetic field, remanent magnetization observed from orbit and in lunar samples indicates that there was a lunar magnetic field in the distant past. Since the magnetic fields of terrestrial planets are thought to be generated in their liquid iron cores, the rotational dynamics of the Moon's core throughout lunar history is directly relevant to studies of ancient lunar magnetism.

Lunar rotational dynamics is very different from that of Earth. The Moon is in synchronous rotation due to Earth's tidal forces and the resulting tidal dissipation, and this state was likely established very soon after the Moon's formation. Tidal dissipation on  Earth over billions of years made the Moon's semimajor axis expand from several Earth radii ($R_E$) at formation to the present value of 60.3 $R_E$. As the lunar orbit and orbital period grew, so did the Moon's rotation period. While the Moon's rotational period is equal to its orbital period, the Moon's obliquity is determined by the so-called Cassini States. Cassini States are minimum-energy solutions for the forced obliquity of a rotating body that experiences tidal dissipation, and most synchronous satellites have evolved into stable Cassini States \citep{col66, pea69}. The spin axis of the satellite in a Cassini State is in the same plane as the normals to the orbital plane and the Laplace plane (the plane around which the orbit precesses). When the period of the axial precession is longer than that of orbital precession (as is the case for the present-day Moon), the satellite is in Cassini State 2, in which the spin axis is closer to the Laplace plane normal than it is to the orbit normal (in the present-day Earth-Moon system, the Laplace plane is the ecliptic). \citet{war75} first established that the Moon underwent a transition between the Cassini States 1 and 2 at a semimajor axis of about 34$R_E$, assuming that the Moon's current shape was already established at this time. The necessity of the Cassini State transition is dictated by the dependence of the lunar spin and orbital precession on Earth-Moon separation. The lunar spin precession is driven by Earth's tidal torque and therefore slowed down with the Moon's tidal recession. However, beyond the Earth-Moon distance of about $15 R_E$ (within which Earth's oblateness was an important perturbation) the tidal recession led to speeding up of lunar apsidal precession, as it is driven by solar perturbations. Therefore, to reach the present state where the natural periods of nodal and spin precession are about 19 and 80~yr, the Moon had to experience equality of these two periods at some point in its evolution, which \citet{war75} placed at the Earth-Moon distance of about $34~R_E$. \citet{war75} also found that the Moon would experience very large forced obliquities during the Cassini State transition. 

\begin{figure}[h]
\centering
\fbox{\includegraphics[scale=.5]{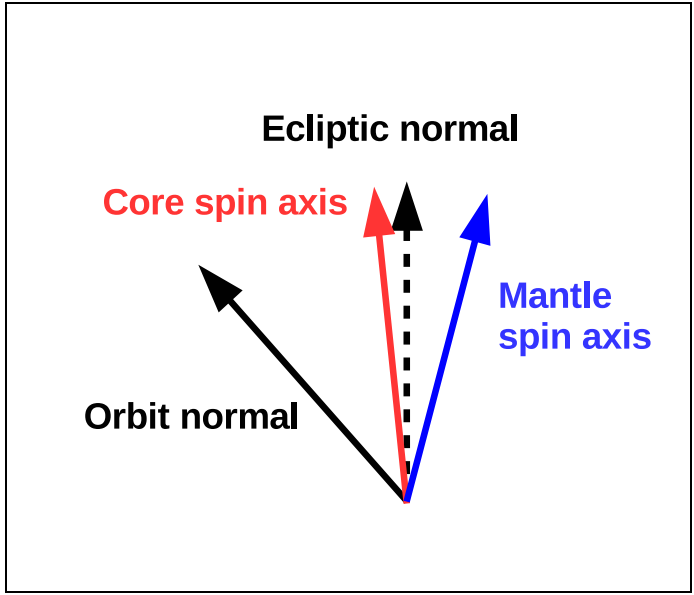}}
\caption{The modern-day relative orientations of the lunar mantle and core spin axes in the plane defined by the ecliptic and lunar orbit normals. The mantle is in Cassini State 2, while the core is in "quasi-Cassini State 2" relative to the mantle. The angles are not to scale: the lunar orbital inclination to the ecliptic is about $5.2^{\circ}$, the mantle obliquity to the ecliptic is about $1.6^{\circ}$ and the core obliquity to the ecliptic is only about $0.1^{\circ}$.}
\label{vectors}
\end{figure}

\citet{mey11} examined how the Moon's liquid core would behave during the obliquity evolution established by \citet{war75}. \citet{mey11} show that the Moon's free core nutation (FCN) period should be several centuries, much longer than the current nodal precession period of the lunar orbit (18.6~yr). Since the Moon's spin axis is in a Cassini State and its precession follows that of the orbit, the natural frequency of nutation of the core around the mantle is too slow to allow the core to follow the mantle's motion. Instead, the lunar core spin axis stays close to the ecliptic normal, in an arrangement somewhat equivalent to Cassini State 2 (the core spin axis is in the plane defined by the ecliptic normal and the mantle spin axis, and the two spin axes are on the opposite sides of the ecliptic normal; Fig. \ref{vectors}). This is in contrast to the case of Earth, where FCN takes only about 400 days, coupling the core to the mantle, which itself is precessing much more slowly, with a 26 kyr period. \citet{mey11} find that this decoupling of the lunar core's rotation axis from that of the mantle should have happened relatively early in lunar orbital evolution, when the Moon had semimajor axis $a=26-29 R_E$, before the mantle's Cassini State transition. Note that the core being ``de-coupled'' from the mantle means that the two are spinning around different axes with different precessional motion; this terminology does not imply that there is no contact or no interaction between the core and the mantle. \citet{dwy11} recognized that the results of \citet{mey11} imply a large past mutual obliquities between the core and the mantle, and \citet{dwy11} argued this differential precession may have been the ultimate driving force behind the lunar dynamo.

More recently, \citet{che16} have shown that lunar obliquity tides should have significantly affected the past lunar inclination. \citet{cuk16} proposed a model in which the Earth has a very fast spin ($<$2.5 hour period) and a high obliquity (~70$^{\circ}$) after lunar formation. Independent of the constraints from lunar inclination, this very fast initial spin of Earth is a requirement by some of the current lunar formation models in order to explain Earth-Moon isotopic similarity \citep{loc18}. As the lunar orbit grows in the \citet{cuk16} model, it encounters an unstable transition between equatorial and ecliptic Laplace planes \citep{tre09, tam13}. The lunar orbit is temporarily trapped in this zone, where solar perturbations induce large eccentricities, loss of the system's angular momentum (which is transferred to the heliocentric orbit), a reduction of Earth's obliquity, and a large lunar inclination ($\simeq 30^{\circ}$). In this picture, the Moon has a high inclination as it enters the Cassini State transition, with strong obliquity tides during the transition reducing the inclination to close to the observed values. In this paper we will consider the dynamics of the lunar core in the context of the \citet{cuk16} model, and determine the consequences for the core's orientation and the Moon's tidal evolution.

\section{Numerical Methods}

In this work we use a custom made numerical integrator {\sc cr-sistem}, which combines mixed-variable symplectic integrator for orbital motion and a Lie-Poisson mapping for the lunar rotation. In all aspects except the treatment of lunar rotation (including the motions of the core), {\sc cr-sistem} is identical to {\sc r-sistem} used by \citet{cuk16}. Therefore we direct the reader to the Methods section of \citet{cuk16} for all details regarding the treatment of mutual perturbation between the bodies, tides within Earth and the Moon, and the shape and precession of Earth. In this section we will describe our approach for integrating the rotations of the lunar mantle and core, and their mutual interaction.

\citet{cuk16} used the approach of \citet{tou94a}, who advance the rotation of the Moon in the body-fixed reference frame implicitly, while the changes to this motion arising from oblateness and triaxiality are integrated explicitly \citep[this algorithm is ultimately based on the Poincar{\'e}-Hough method][]{hou95, poi10}. The fact that these perturbations to the uniform rotation are relatively small enabled \citet{cuk16} to efficiently integrate the rotational dynamics with only dozens of time steps required for each orbital/rotational period. However, inclusion of a triaxial core within a triaxial mantle makes the Hamiltonian much more complex. 

\citet{tou01} discuss in some detail the optimal algorithms for numerically integrating the rotational dynamics of a planet with a solid mantle and a liquid core. While \citet{tou01} are able to derive a reasonably efficient algorithm for the dynamics of an oblate (i.e. rotationally symmetric) planet with a core, there is no such option for a triaxial core-mantle system. As suggested by \citet{tou01}, we divide our Hamiltonian into three parts, which correspond to rotations around the three Cartesian axes:
\begin{eqnarray}
\mathcal{H}_x = {1 \over 2 \alpha} \Bigl(A_C P^2 + A P_C^2 - 2 F_C P P_C \Bigr)\\
\mathcal{H}_y = {1 \over 2 \beta } \Bigl(B_C Q^2 + B Q_C^2 - 2 G_C Q Q_C \Bigr)\\
\mathcal{H}_z = {1 \over 2 \gamma} \bigl(C_C R^2 + C R_C^2 - 2 H_C R R_C \Bigr)
\label{hamil}
\end{eqnarray}
where $A, B, C$ are the principal moments of inertia and $P, Q, R$ are the components of angular momentum measured with respect to the mantle's principal axes (values without subscript refer to the whole body, and those with the subscript ''C'' to the core. As \citet{tou01} show, $P_C$, $Q_C$ and $R_C$ are not really the angular momentum components of the core, but function equivalently. Additionally, the core moments of inertia are $F_C={2 \over 5}m_C b c$, $G_C={2 \over 5}m_C a c$, $H_C={2 \over 5}m_C a b$, and we define $\alpha=A A_C-F_C^2$, $\beta=B B_C-G_C^2$ and $\gamma= C C_C-H_C^2$, where $m_C$ is the core's mass and $a$, $b$ and $c$ are the core's principal axes \citep[our approach and notation follows that of ][]{tou01}. These three Hamiltonians each generate separate rotations of total and core angular momentum vectors ${\bf M}$ and ${{\bf M}_C}$ around the corresponding mantle axes. For example, the angular velocities generated by $\mathcal{H}_z$ for ${\bf M}$ and ${\bf M}_C$ are $w_z=(C_CR-H_CR_C)/\gamma$ and $w_{Cz}=(CR_C-H_CR)/\gamma$ (with the sense of rotations reversed for $w$ and $w_C$ due to our mantle-fixed reference frame). Using these angular velocities, we can define rotation matrices that convert vectors between the mantle-fixed and inertial frames, and calculate periodic ``kicks" on the rotational motion due to gravitational torques and tidal dissipation. 

It is clear from our definition of $w$ that the rotation frequencies of the mantle around all three axis include terms independent of the core's properties (at least to the lowest order) and therefore $w$ does not vanish even in the limit of the core's size and mass being zero. The core-mantle cross-terms do become negligible, but the main components of $\dot{\bf M}$ remain, and correspond to the rotation of ${\bf M}$ around the mantle (or, in an inertial frame, rotation of the mantle around the total angular momentum vector). This means that the integrator needs to resolve this basic rotational motion which is handled explicitly, in contrast to the treatment in \citet{tou94a} \citep[and subsequently ][]{cuk16} where only the slower nutational motion (separate from principal axis rotation) had to be resolved. Our numerical model which includes the lunar core therefore requires significantly larger number of time-steps per rotation period. There is a direct parallel here to orbital integrations, with our integrator being analogous to a basic $T+U$ symplectic integrator \citep{for90}, which requires enough time-steps to resolve basic Keplerian motion, while the \citet{tou94a} scheme is analogous to mixed-variable symplectic integrators \citep{wis91} which assume Keplerian orbits, so fewer time-steps are needed per orbit. Therefore, we are not in a good position to replicate the long-term integrations of \citet{cuk16}, as inclusion of the lunar core would increase integration time by at least an order of magnitude. Therefore we will restrict ourselves to examining how the presence of the lunar core affects the Cassini States of the Moon at different Earth-Moon distances, and determining the mutual motions of the core and the mantle for the purposes of understanding the past lunar dynamo.

One remaining choice we had to make was the introduction of core-mantle friction. We calculate the vector difference between the rotational velocities of the mantle and the core $\Delta {\vec \omega}$, and then apply a torque to $\bf{M_C}$ every time-step:
\begin{equation}
d{\bf M}=- C_C \Delta {\vec \omega} K_{CM} dt  
\label{friction}
\end{equation}
where $dt$ is the time-step and $K_{CM}=8.43 \times 10^{-3}$ year$^{-1}$ is the magnitude of spin-orbit friction. We based this form of core-mantle torque that is linear in the core-mantle relative motion on Eq. 19 in \citet{pav16}. Our value for $K_{CM}$ corresponds to a damping timescale of $(1000 / 2 \pi)$ sidereal months, or 120 years, and is equivalent to \citet{pav16} $k_v/C_T=16 \times 10^{-9}$~day$^{-1}$. Our use of linear core-mantle boundary (CMB) friction follows the usage in LLR community, but the flow at the CMB boundary is likely to be turbulent and non-linear \citep{wil01}, so our model should be taken as a first approximation. As we ignored any tidal forces from Earth or the Sun on the core, the core-mantle friction term (Eq. \ref{friction}) is the only dissipative influence on the core's rotation, and it acts to match it to the rotation of the mantle. However, non-dissipative core-mantle torques inherent in our integrator make the core precess at a different angle and phase from the mantle, which can lead to equilibrium points very different from core-mantle co-rotation, as we show in the following sections.

\section{Core-Mantle (De)coupling}

After completing our integrator, the first situation we wanted to explore is the present-day dynamics of the lunar core. We used the same lunar shape parameters as \citet{cuk16}, and we decided to set the mass and radius of the lunar core exactly to 0.01~$M_M$ and 0.2$R_M$, and we will use these values throughout the paper. Note that the ellipticity of the core is by far the most important parameter for its dynamics, and the mass and density have only second-order effects on the core's nutation. Here we used shape parameters $(C_C-A_C)/C_C=2.5 \times 10^{-4}$ and $(C_C-B_C)/C_C=2 \times 10^{-4}$ for the core \citep[ellipticity was based on ][]{pav16}.  We used long-term average lunar tides (quantified by tidal quality factor $Q_M=60$ and Love number $k_{2M}=0.024$). 

\begin{figure}[h]
\centering
\includegraphics{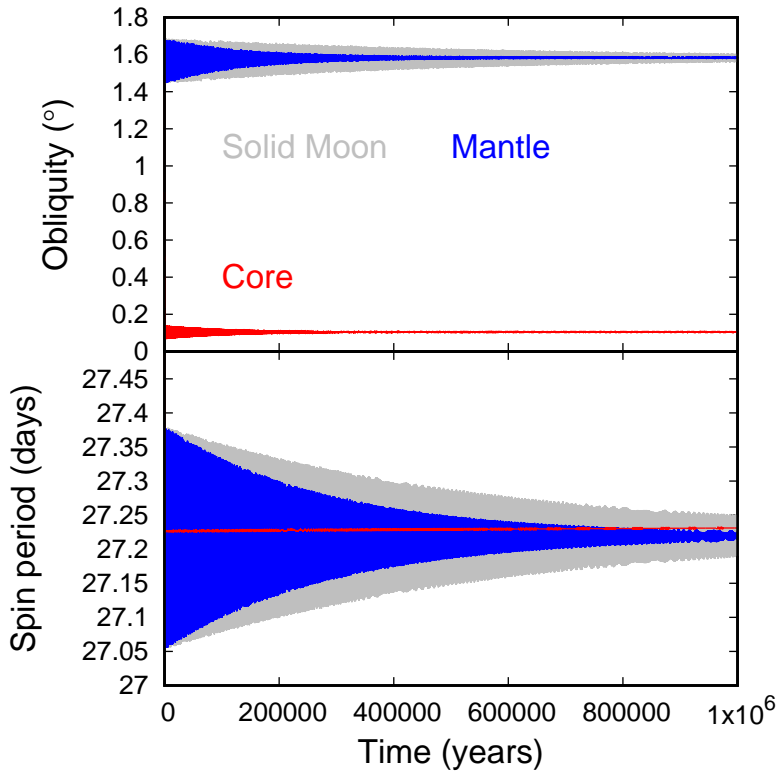}
\caption{Top: Evolution of the ecliptic obliquities of the lunar core (red) and the mantle (blue), as well as a solid moon (gray, mostly covered by blue) in the present Earth-Moon system. Bottom: Evolution of the spin periods of the solid Moon (gray), lunar mantle (blue) and the core (red) in the same simulations (pattern of waves in blue and black curves are an artifact of the output frequency, the real libration period is much shorter). Initial conditions had the core slightly sub-synchronous and with a spin axes aligned with that of the mantle.}
\label{nom1}
\end{figure}

Figure \ref{nom1} shows a $10^6$~yr simulation in which the mantle was put in synchronous rotation and Cassini State 2 with obliquity to ecliptic $\epsilon_M=1.5^{\circ}$, while the core is made to approximately co-rotate with the mantle. For comparison, we also conducted a separate experiment using an entirely solid Moon with no fluid core. In about 300 years the core decouples from the mantle's spin direction and settles in an apparently stable state with obliquity to ecliptic of $\epsilon_C=0.1^{\circ}$. The spin rate of the core on the other hand settles close to that of the mantle, firmly within the range of mantle's rotational librations (bottom). The mantle is meanwhile damping librations around its Cassini state close to $1.58^{\circ}$, with the damping timescale being much faster than for a solid moon, indicating that the damping is dominated by CMB friction.   

\begin{figure}[h]
\centering
\includegraphics{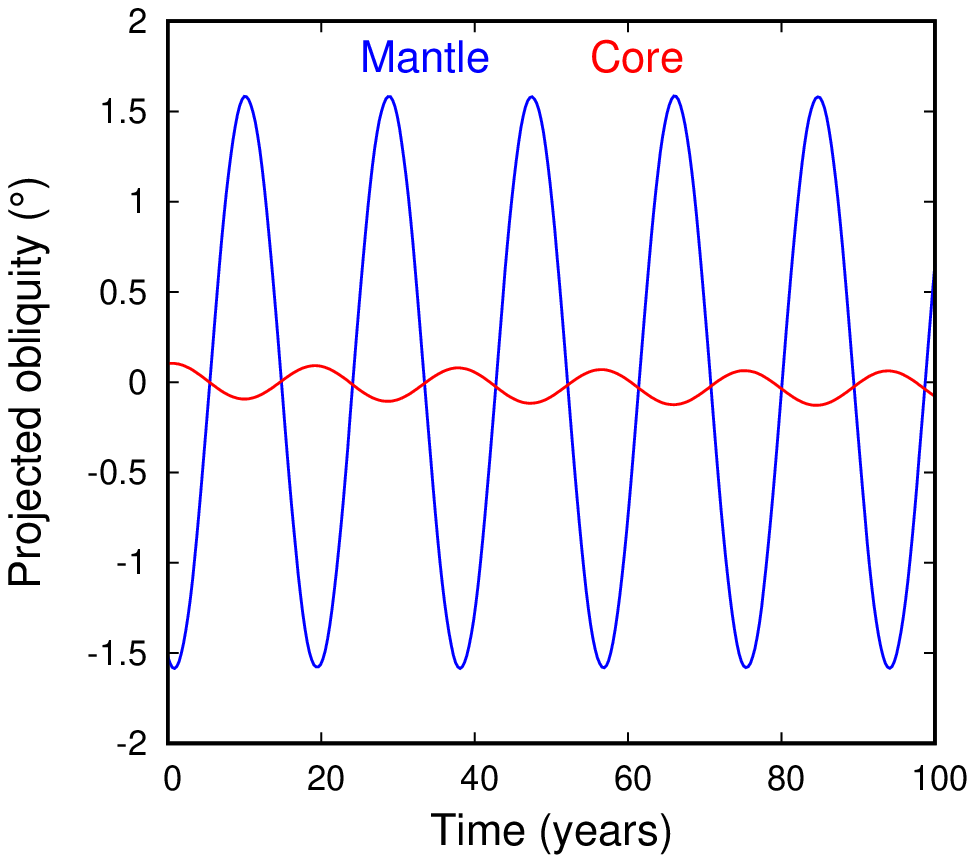}
\caption{The very end of the simulation shown in Fig. \ref{nom1}. The lines plot a projection of the obliquity of the core (red) and mantle (blue) on the $y$-axis in a non-rotating 2-D Cartesian frame centered on the ecliptic pole. It is clear that the mantle and the core spin axes precess around the ecliptic normal with opposite phases, which is analogous to core being in a quasi-Cassini State 2. }
\label{nom2}
\end{figure}

Figure \ref{nom2} shows the situation at the end of the same simulation as shown in Fig. \ref{nom1}. Here we plot the $y$-component the core and mantle spin vectors in the polar representation $\epsilon_i \cos(\Omega_i)$, where $\epsilon$ is the obliquity with respect to ecliptic and $\Omega$ the longitude of the node (i.e. intersection between the plane normal to the spin axis and the ecliptic) for either the mantle and the core. This plot enables us to see how the obliquity vectors are oriented in space. From Fig. \ref{nom2}, it is clear that the mantle and the core spins are both precessing around the ecliptic normal with the 18.6~yr period dictated by the (orbital) nodal precession. While the core's obliquity is much smaller, it is precessing exactly out-of-phase (i.e. $180^{\circ}$ away) from the mantle, as expected from a "quasi-Cassini State 2" (see Fig. \ref{vectors}) in which the natural precession frequency of the core (around the mantle) is much slower than the mantle's precession around the reference direction (in this case, the ecliptic normal). 
As explained in \citet{mey11}, the core should precess around the mantle with the frequency:
\begin{equation}
\omega_{FCN}=\omega \Bigl({C_C-A_C \over C_C}\Bigr) \Bigl({C  \over C_M}\Bigr)
\label{nutation}
\end{equation} 
where $\omega$ is the lunar spin rate, and $C_M=C-C_C$ the mantle moment of inertia (for our purposes, $C/C_M \simeq 1$). Since we use $(C_C-A_C)/(C_C)=2.5 \times 10^{-4}$, and the lunar spin period is 27.3~days, then the period for the core nutation should be $\simeq 300$~yr. Since $\epsilon_C << \epsilon_M$, we can estimate $\epsilon_C/\epsilon_M \simeq \omega_{FCN} / \dot{\Omega}$, where $\dot{\Omega}$ is the nodal precession rate for the Moon \citep[based with analogy with Cassini State 2][]{war75}. Therefore, the expected forced obliquity of the lunar core should be $\epsilon_C = 1.6^{\circ} (18.6 {\rm yr} / 300 {\rm yr}) \simeq 0.1^{\circ}$, fully consistent with the results of our simulation. These results are also very similar to the findings of \citet{dum16} and \citet{sty18} who additionally took into account the existence of the inner solid core. So we can conclude that for the present Earth-Moon system, our numerical model returns an answer consistent with the analytical expectations, with the lunar core precessing separately from the mantle with a much smaller obliquity to the ecliptic.

\begin{figure}[h]
\centering
\includegraphics{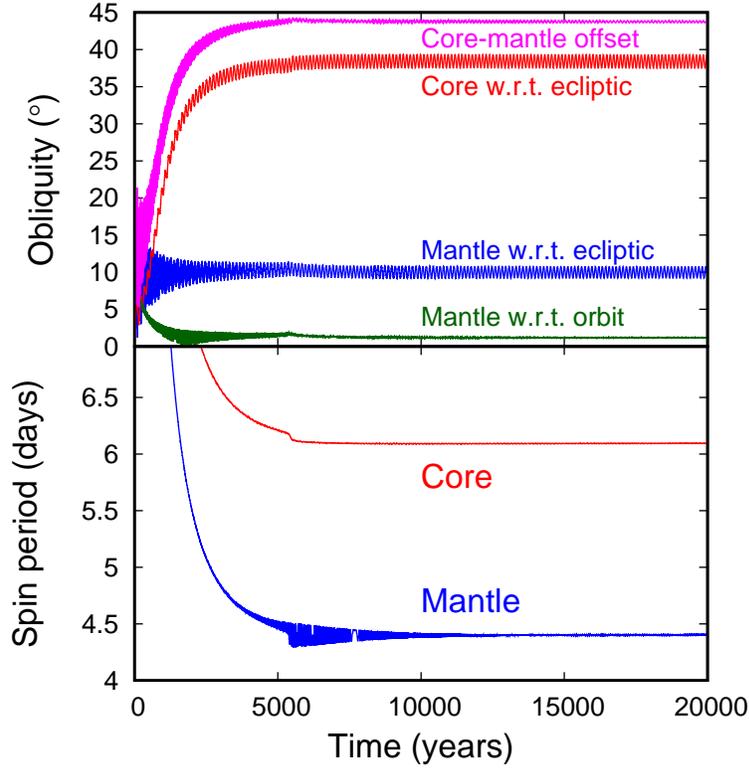}
\caption{Simulation of relaxation to equilibrium state using lunar semimajor axis $a=18 R_E$, eccentricity $e=0$, and inclination $i=8^{\circ}$, with core flattening $f=2.5 \times 10^{-4}$ . The core is largely decoupled from the mantle in this simulation. Top: Evolution of the ecliptic obliquities of the lunar core (red) and the mantle (blue and green, relative to ecliptic and the lunar orbit, respectively), as well as their relative offset (magenta). Bottom: Evolution of the spin periods of the solid Moon (black), lunar mantle (blue) and the core (red) in the same simulation.}
\label{test1}
\end{figure}

\begin{figure}[h]
\centering
\includegraphics{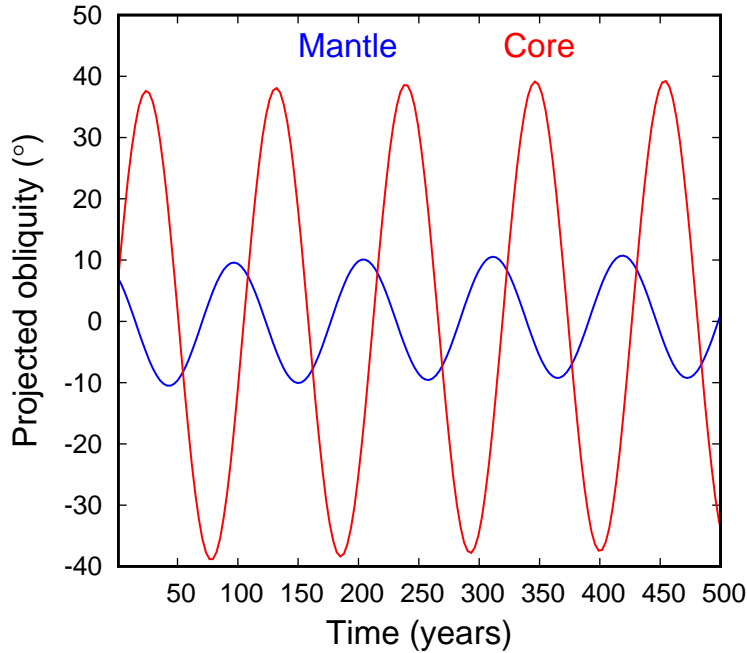}
\caption{The very end of the simulation shown in Fig. \ref{test1}. The lines plot a projection of the obliquity vectors of the core (red) and mantle (blue) on one of the axis in the polar representation. It is clear that the mantle and the core spin axes precess around the ecliptic normal with very different phases, indicating dynamical decoupling between the core and the mantle. }
\label{test1r}
\end{figure}

\begin{figure}[h]
\centering
\includegraphics{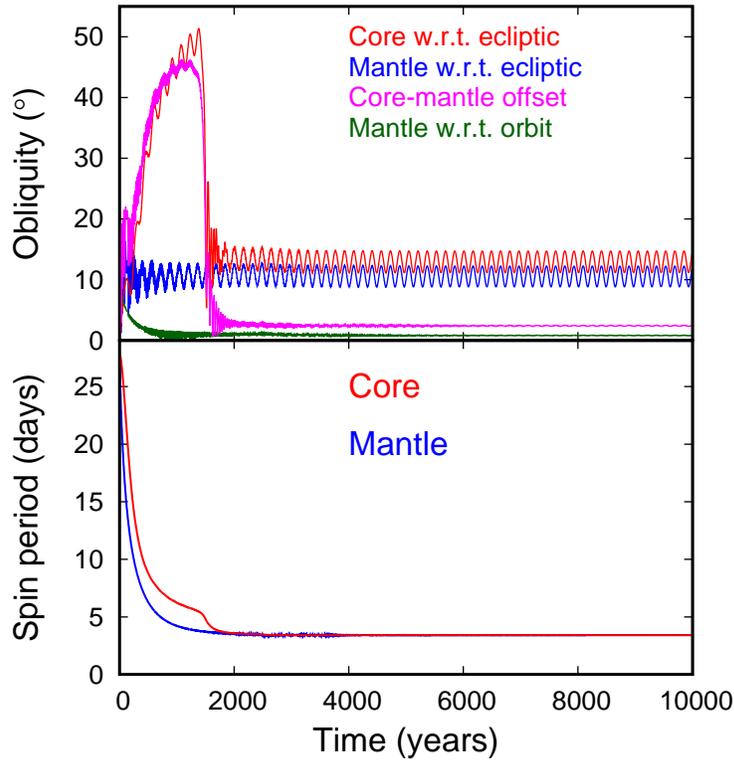}
\caption{Simulation of relaxation to equilibrium state using $a=15 R_E , e=0, i=8^{\circ}$, with core flattening $f=2.5 \times 10^{-4}$. Free core nutation is fast enough to couple the core to the mantle. Top: Evolution of the ecliptic obliquities of the lunar core (red) and the mantle (blue and green, relative to ecliptic and the lunar orbit, respectively), as well as their relative offset (magenta). Bottom: Evolution of the spin periods of the lunar mantle (blue) and the core (red) in the same simulation.}
\label{test2}
\end{figure}

\begin{figure}[h]
\centering
\includegraphics{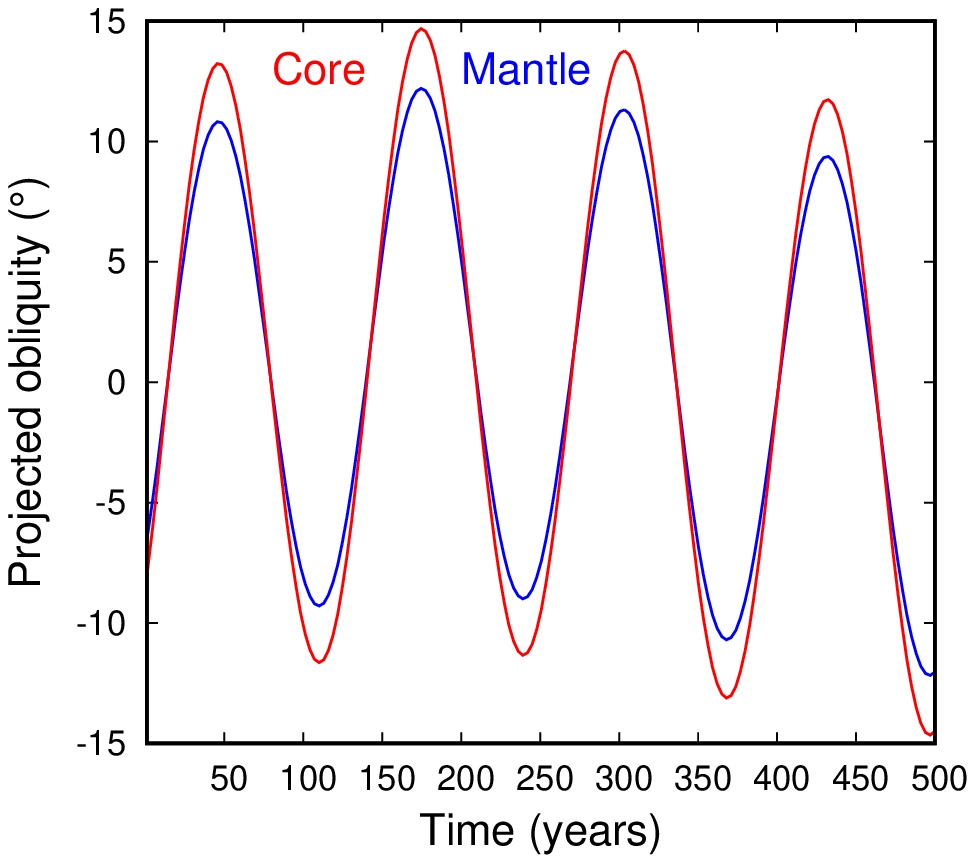}
\caption{The very end of the simulation shown in Fig. \ref{test2}. The lines plot a projection of the obliquity vectors of the core (red) and mantle (blue) on one of the axis in the polar representation. It is clear that the mantle and the core spin axes precess around the ecliptic normal in-phase, which indicates that the core is in quasi-Cassini State 1. }
\label{test2r}
\end{figure}

We now turn our attention to the transition between the core being coupled and uncoupled from the mantle, which is in many ways analogous to the Cassini State transition that the mantle experienced somewhat later, with the difference that the former involved the core and the mantle, and the latter the mantle and the lunar orbit. \citet{mey11} calculated that the core should have uncoupled at around $26 R_E$, assuming that the mantle was not in hydrostatic equilibrium but had a figure identical to the present one. As explained in \citet{cuk16}, this is probably the smallest Earth-Moon distance at which a non-hydrostatic mantle would be a reasonable assumption. However, in our simulations of the Moon at this distance from Earth, we did not exactly reproduce \citet{mey11} result, and we kept getting an uncoupled core in quasi-Cassini State 2 (note that the lunar mantle is at this point in Cassini State 1 relative to the orbit). We tried simulations at even smaller Earth-Moon distances, but even for $a=18 R_E$ we were getting a decoupled core (Figs. \ref{test1} and \ref{test1r}), using $(C_C-A_C)/C_C=2.5 \times 10^{-4}$ (as elsewhere in the paper). However, when we reduced the semimajor axis to $a=15 R_E$, the core in the simulation is now coupled to the mantle (Figs. \ref{test2} and \ref{test2r}). Note that these simulations only intended to demonstrate that the core-mantle coupling is possible for small Earth-Moon distances, and are not in any way an accurate representation of the Moon's dynamics at the time, as the mantle was likely not rigid at this time, and we did not try to accurately match Earth's rotation and oblateness which affect the Moon's orbital precession.

We conclude that the criterion used by \citet{mey11} for the core-mantle coupling, $\omega_C > \dot{\Omega}$, is approximate and cannot be used to accurately determine the distance at which the lunar core dynamically decoupled from the mantle. Numerical integrations suggest that the decoupling probably happened at lunar semimajor axis somewhat smaller than $26 R_E$ estimated by \citet{mey11}. This is a relatively minor discrepancy, as the nutation period of the core is about 50\% and 33\% of the mantle precession period in our Figs. \ref{test1} and \ref{test2}, respectively. Therefore, our simulations suggest that dynamical decoupling happens when the core nutation rate is about twice the nodal precession rate. A more accurate determination would require many more simulations and is of limited interest to this paper. \citet{dum16} have conducted an analysis of core-mantle coupling (but without dissipation) and found that the exact identity of the mantle precession rate and the FCN is in fact a resonance which forces a large core-mantle tilt; they find that FCN must be significantly faster than mantle precession to enable dynamical coupling. Therefore, the main conclusion of this section is that, assuming a rigid lunar mantle with the present shape, the core should be dynamically decoupled from the mantle for lunar semimajor axes $a > 25 R_E$ that are relevant for the Cassini State Transition and lunar paleomagnetism.

\section{Effects of the Core on the Lunar Dynamical Evolution}

In order to have a self-consistent model of past lunar core dynamics, we need to re-examine the dynamical history of the lunar mantle with the core's influence included. Lunar inclination damping due to obliquity tides, as modeled by \citet{cuk16}, directly depends on the mantle's rotational state, which is in principle affected by the presence of the core. It is important to revisit these calculations and determine whether the results obtained by modeling the Moon as a solid body are still valid when core-mantle interaction is included. However, due to the limitations of our current integrator relative to that used by \citet{cuk16}, we are not able to directly integrate the lunar tidal evolution, as complete simulations would take unreasonably long time. 

In this section we will instead present a series of simulations of relaxation to equilibrium of the lunar core and mantle. These simulations are designed to follow the evolutionary track in semimajor axis and inclination found by \citet{cuk16}. We started these simulations with a sub-synchronous Moon, let it damp to a stable state, and recorded the final rotational state of the core and the mantle. While this is not a model of lunar orbital history independent of \citet{cuk16}, is an important test of their model. If the rotational state of the mantle, especially the obliquity, is fundamentally changed by the presence of the core, the results of \citet{cuk16} would not be valid. On the other hand, if the simulations with the core-mantle interactions included produce lunar obliquities close to those in the core-less model, the overall nature of the lunar tidal evolution is likely unchanged by core-mantle interactions. 

\begin{figure}[h!]
\centering
\includegraphics{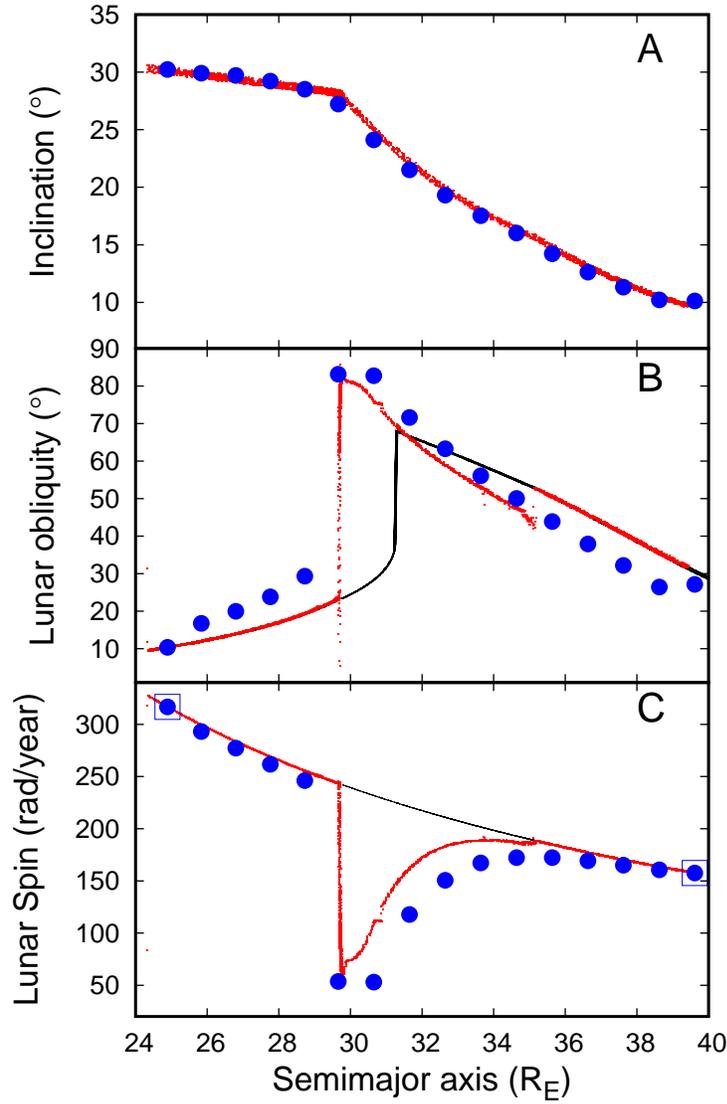}
\caption{Equilibrium solutions for the mantle spin state using our core-mantle code (solid blue circles) along the lunar tidal evolution track found by \citet{cuk16} (small red dots). Panel A plots the lunar orbital inclination as a function of semimajor axis, while the panels B and C plot the corresponding lunar (i.e. mantle) obliquity (with respect to lunar orbit) and spin rate, respectively. The thin black lines in panels B and C plot the obliquity and spin rate calculated on the assumption of the Moon being in synchronous rotation and Cassini State while on the $a$-$i$ track shown in panel A. The open blue squares in panel C indicate that the lunar mantle was in synchronous rotation at the end of the simulation.} 
%The comparison between the \citet{cuk16} results for a core-less Moon and our results for the full core-mantle system shows a great degree of convergence, with noticeable divergence happening only for semimajor axes immediately interior to the Cassini State transition (27-29 $R_E$).}
\label{cassini}
\end{figure}

In Fig. \ref{cassini}, we plot the results of our equilibrium solutions along the track of the lunar tidal evolution found by \citet{cuk16}. We generated these solutions by placing the Moon at a sequence of semimajor axes, with the spacing of 1 $R_E$. At each semimajor axis, the Moon was given orbital inclination based on the orbital evolution track taken from Figure 4 in \citet{cuk16} (the same data are plotted by small red dots in Fig. \ref{cassini}), while the eccentricity was set to $e=0.01$. The solid blue circles show the end-states of the simulations after 1 Myr (extended to 2 Myr for the two simulations at 30-31 $R_E$ which showed slowest convergence) averaged over the final 1000 years. Due to short-period variations in orbital elements, our initial conditions and averaged final states do not perfectly match in either inclination or semimajor axis, causing the blue circles to be slightly misaligned with the red dots they track.

In panel B, we compare our solutions for the mantle obliquity for the specific $a$ and $i$ plotted in panel A (solid blue circles) to the core-less Moon's obliquity found by \citet{cuk16} (small red dots). The solid black line plots the Cassini State obliquities corresponding to the inclinations in panel A (unlike the numerical simulations, the Cassini State calculation, by definition, assumes synchronous rotation). In panel C we plot the corresponding rotational rates from our mantle equilibrium solutions (solid blue circles) and the \citet{cuk16} core-less Moon simulation (small red dots), with the thin black line plotting the synchronous state.

Our solutions clearly show similarity to the previously found evolution, including the divergence from a Cassini State. However, while the divergence from the Cassini state for a core-less moon was limited to the immediate aftermath of the Cassini State Transition, our simulations suggest that the Moon was in non-synchronous rotation for all semimajor axes in the $a=25-40 R_E$ range. Particularly interesting is the divergence from \citet{cuk16} results in semimajor axis range 26-29 $R_E$, where our mantle is in a non-synchronous state, while \citet{cuk16} find the Moon to be in synchronous rotation. There are two distinct reasons for this divergence. One is from the different ways the two solutions were calculated: \citet{cuk16} integrated a single continuous simulation of lunar tidal evolution, while we performed a number of short simulations that started with the Moon in non-synchronous rotation. Therefore, in this segment of \citet{cuk16} simulations the Moon is likely in a quasi-stable Cassini State 1, which is temporarily stable against Earth's torques, but would not be re-established if it were to be broken by an outside factor (such as a large impact). In the \citet{cuk16} simulation, the synchronous rotation is broken by an annual libration resonance at 29.7 $R_E$, after which the Moon settles in a sub-synchronous state adjacent to Cassini State 2 (their Extended Data Figs. 8 and 9). In our core-mantle simulations the Moon at $a=27-29 R_E$ is in a sub-synchronous state adjacent to Cassini State 1 (Fig. \ref{cassini}), which apparently does not exist in the \citet{cuk16} model (their Extended Data Figs. 7 and 8).

% While we do not have space here to discuss the dynamics of this new state, we note that the existence of this sub-synchronous quasi-Cassini State 1 makes the lunar tidal evolution (and especially inclination damping) less dependent on stochastic events such as impacts. If the Moon is ejected out of Cassini State 1 by an impact while at $a=27-29 R_E$, we find that it would settle in a relatively similar non-synchronous state, and not in a diametrically opposite quasi-Cassini State 2, which would dictate an extremely high obliquity. 

Apart from the divergence prior to the Cassini State transition, we also observe that our simulations at semimajor axes 35-40 $R_E$ have the Moon in a non-synchronous state, while the \citet{cuk16} find the Moon in the synchronous Cassini State 2 at this distance (beyond 40 $R_E$ we also find that the Moon is mostly in synchronous rotation). The main reason for the prevalence of non-synchronous rotation in our simulations is that the core, which lags behind the mantle in its spin rate, exerts a relatively substantial spin-changing torque on the mantle. This can move the equilibrium spin value outside of the synchronous spin-orbit resonance, producing a stable sub-synchronous state. While for some semimajor axes $a=30-35 R_E$ the synchronous state itself may be unstable \citep[due to too-high Cassini State obliquity;][]{bel72}, we find that in the $a=35-40 R_E$ range the synchronous rotation is sometimes reachable by spin-down but not by spin-up (i.e. initial conditions may determine the end point rotation state reached by the Moon).     

In general, the presence of the core drives the sub-synchronous states further away from synchronicity, and as a result the corresponding Cassini States have somewhat higher obliquities. This can be understood by recognizing that the pole of a slower-rotating Moon will precess more quickly, so to match the precession rate of the orbit, axial precession must be slowed down by high obliquity. On the other hand, we find that the Cassini State obliquities for synchronous cases (such as our $a=40 R_E$ simulation) are lower when the core is included in the simulation. The reason for this change is that the core, which is significantly misaligned with the mantle, induces additional librations of the mantle in synchronous rotation. Librations make the precession slower (as the Moon presents a different average figure to Earth's torques), requiring lower obliquities to reach the same precession rate as in the non-librating case. 
 
Despite the modification of synchronous rotation caused by the presence of the fluid core, we find that the lunar orbital evolution history found by \citet{cuk16} is unlikely to be fundamentally changed. Using the lunar inclination model published by \citet{cuk16}, we find that the lunar mantle has had a large obliquity for the significant part of the Moon's history. Obliquities and associated tides in our model are sometimes lower, sometimes higher than those found by \citet{cuk16}, but there is every reason to think that large damping of lunar eccentricity took place. We also note that core-mantle friction also damps obliquity relative to the Laplace plane \citep{roc76}, which leads to damping of inclination (as the obliquity is forced). Further more extensive numerical simulations of the full evolution of lunar spin and orbit in the presence of the core are needed to determine the exact contributions of different inclination-damping mechanisms.

Our results are dependent on our choice for the functional form of CMB friction (Eq. \ref{friction}), in which the torque is directly proportional to the difference between core and mantle spin rates, with a constant damping timescale of about 120~yr. Other functional forms are possible, including those that are quadratic in relative core-mantle motion \citep{yod81}. Given that the current core-mantle tilt is small, these alternative forms would produce much shorter timescales for damping of core-mantle motion when the lunar obliquity to ecliptic was high. We tried such a functional form of CMB friction in some simulations, in order to determine whether such strong friction could couple core to the mantle. Surprisingly, we find that beyond about $30 R_E$ the CMB torque on the mantle is too strong for mantle to stay in the Cassini State, and the Moon evolves into a very slow-rotating, high-obliquity state with the core and the mantle still decoupled. We expect that this state would persist until the lunar inclination is damped, ultimately allowing core-mantle coupling in the absence of mantle's forced obliquity. As a substantial lunar inclination has survived the Cassini State transition, this puts an upper limit on the intensity of past CMB friction, and may indicate that past quadratic CMB friction is not consistent with the present lunar orbit. However, our model includes a number of approximations, and we only explored relatively short-term dynamics of the core and the mantle at fixed Earth-Moon distances; full-scale evolutionary integrations are necessary to fully constrain long-term inclination damping due to quadratic CMB friction.

\section{Past Rotational State of the Core}

\begin{figure}[h]
\centering
\includegraphics{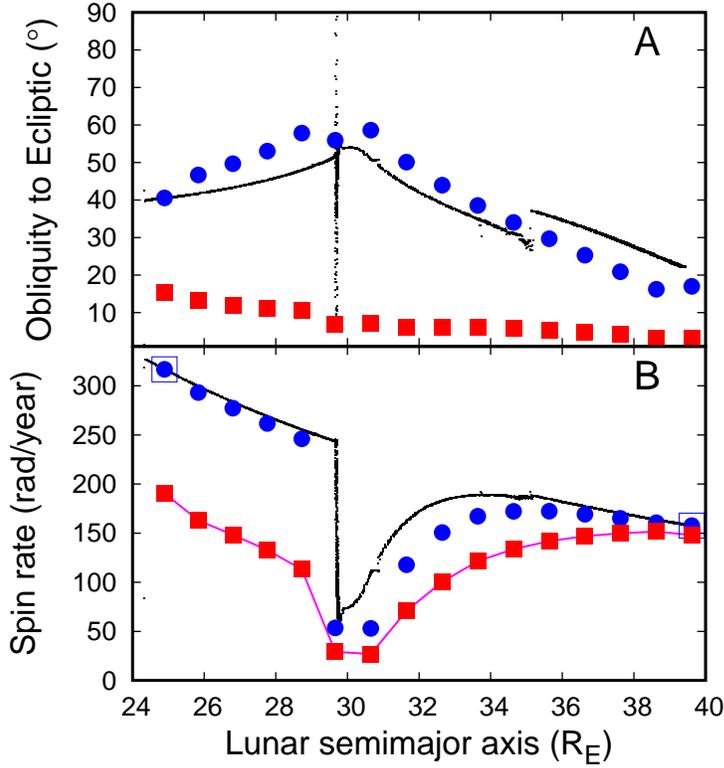}
\caption{Comparisons between the rotational parameters of the lunar mantle and the core in the simulations shown in Fig. \ref{cassini}. Obliquity to ecliptic is shown in the top panel, while the bottom panel plots the core and mantle rotation rates. Obliquity and spin rate of the mantle (solid blue circles) and the core (solid red squares) are determined by our full core-mantle simulations using {\sc cr-sistem}. The solid-Moon integration of \citet{cuk16} is plotted with small black dots. In the bottom panel, thin magenta line plots the rotation rate $\omega_M \cos(\psi_{CM})$, where $\omega_M$ is the mantle spin rate, and $\psi_{CM}$ is the core-mantle obliquity.}
\label{core1}
\end{figure}

For the purposes of the generation of the lunar magnetic field, we are chiefly interested in the rotational state of the liquid core. In Fig. \ref{core1} we plot the final rotational parameters of the core in the simulations shown in Fig. \ref{cassini}. The solid squares and circles plot the results of our simulations featuring the core, while small dots plot the \citet{cuk16} simulations of a core-less Moon. In the top panel we plot the core and mantle obliquities relative to the ecliptic, a convenient system of reference because the core's spin axis is close to the ecliptic for the duration of the simulations. Unlike the core, the mantle has a much larger tilt to the ecliptic, both due to large orbital inclination (most important before the Cassini State transition) and the large forced obliquity (dominant after the Cassini State transition). As the core is in a non-synchronous rotation, we also plot the predicted core rotation rate based on that of the mantle using the solid magenta line. This rate is calculated as $\omega_C=\omega_M \cos(\psi_{CM})$, where $\omega_C$ and $\omega_M$ are the core and mantle spin rates, and $\psi_{CM}$ is the core-mantle obliquity. This calculated spin rate assumes that the core spin matches the component of the mantle's angular velocity parallel to the core's spin axis. Comparison between these predicted and numerically simulated spin rates of the core (Fig. \ref{core1}, bottom panel) shows this to be a good approximation for the core's final spin rate in the simulations.

\begin{figure}[h]
\centering
\includegraphics{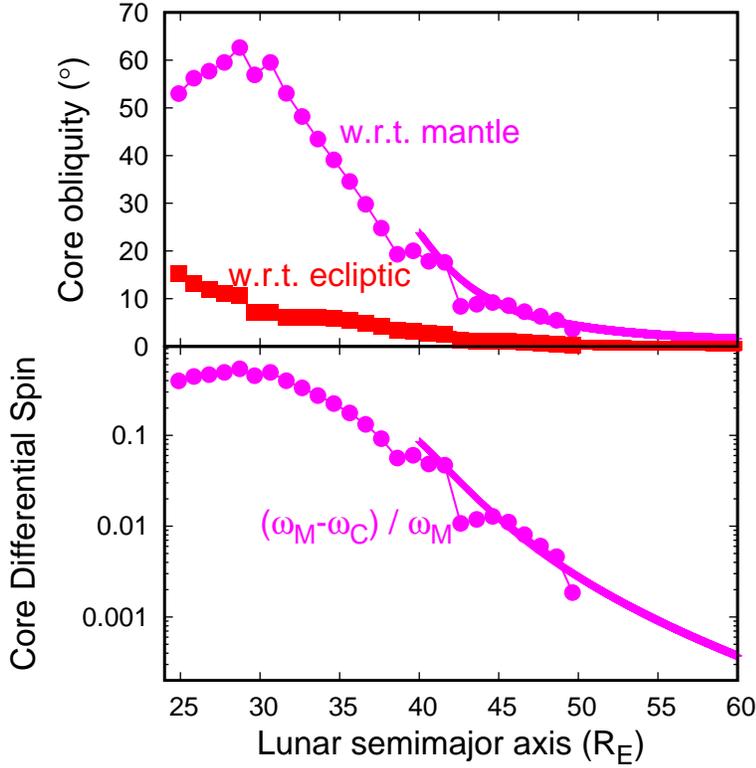}
\caption{Lunar core obliquity (top panel) and relative spin rate (bottom panel) in our model. The points for $a < 50 R_{E}$ were determined by a full numerical model also shown in Figs. \ref{cassini} and \ref{core1}, while the solid line for $a > 40 R_E$ was calculated using Eq. \ref{hybrid} and the semi-analytical model described in the text. In the top panel, the upper magenta lines plot the core-mantle obliquity, while the lower red lines plot the core's obliquity relative to the ecliptic. In the bottom panel, we plot the quantity $(\omega_M-\omega_C)/\omega_M$ as a measure of relative difference between the spin rates of the core and the mantle.}
\label{core2}
\end{figure}
Figure \ref{core2} re-plots some of our results for the core in terms of quantities measured relative to the mantle, and extends the calculation to the present day using a semi-analytical model. Since the Moon was most likely in Cassini State 2 for $a > 40 R_E$, and direct simulations take much longer for large semimajor axes, it is more efficient to calculate core and mantle rotational properties using a simplified model. We find the core obliquity as the solution to the equation:
\begin{equation}
\Bigl({C_C - A_C \over C_C}\Bigr) \omega_M \cos(\psi_{CM}) \sin(\psi_{CM}) + \dot{\Omega} \sin(\psi_{CM}-\epsilon_M) = 0
\label{hybrid}
\end{equation}
where $\dot{\Omega}$ is the precession rate of the longitude of the Moon's ascending node, and $\epsilon_M$ is the obliquity of the mantle with respect to the ecliptic. We obtain Eq. \ref{hybrid} from Eq. \ref{nutation} (in this paper) and Eq. 1 in \citet{war75}, and assumed $\omega_C=\omega_M \cos(\psi_{CM})$ for the core's rotation rate. Eq. \ref{hybrid} can alternatively be derived from the model of \citet{sty18}, by setting the size of the inner core to zero. Eq. \ref{hybrid} is solved numerically by simply advancing through obliquities and finding the value for which the sum of terms goes through zero. The mantle obliquity and lunar orbital precession are obtained from the semi analytical model of lunar tidal evolution described by Eqs. M1-M8 in the Methods section of \citet{cuk16}. We used $i=10^{\circ}$ and $e=0.01$ at $a=40 R_E$ as initial conditions, and advanced the model forward, with Earth tidal $Q$ set at lifetime average of $Q_E=35$, and the lunar tidal $Q$, as well as Earth and Moon tidal Love numbers, set at present values $Q_M=38$, $k_2^E=0.3$ and $k_2^M=0.024$ \citep{wil15}. This model evolution gives us an inclination just over $5^{\circ}$ at $a=60 R_E$, consistent with the present lunar orbit. Thick lines in Fig. \ref{core2} plot the core obliquity and spin rate derived from this model.

Fig. \ref{core2} shows that, while the numerical and semi-analytical obliquities mostly agree, there is a dramatic divergence for points at $a=43 R_E$ and $a=50 R_E$, for which the numerical model does not predict synchronous rotation. This is related to the shifting of the equilibrium spin rate from exact synchronicity due to the CMB friction torque on the mantle. It appears that the equilibrium rotation is just at the outer limit of the synchronous spin-orbit resonance, making reestablishment of synchronous rotation sensitive to initial conditions and possible secondary resonances. Here we used a low lunar eccentricity ($e=0.01$), but a more substantial eccentricity would shift the equilibrium spin value to higher values, counteracting the effects of CMB friction and shifting the equilibrium back into the synchronous rotation. Lunar eccentricity was likely affected by planetary orbital resonances at $a=46 R_E$ and $a=53 R_E$ \citep{cuk07}, with the latter resonance likely bringing the lunar eccentricity close to the present value. The actual history of the lunar rotation depends on the timing of synchronous-lock-breaking impacts, the relative strength of satellite tides and spin-orbit friction, the time-evolution of lunar eccentricity and the (poorly understood but likely) influence of secondary resonances 

It is evident from Fig. \ref{core2} that, in our model, when the Moon had semimajor axis $a < 45 R_E$ there was a large mutual obliquity between the lunar core and the mantle, and a significant difference between the rotation rates of the two. In order to compare the models of orbital and rotational dynamics to the record of lunar magnetism, we would need to know the absolute timing of the lunar tidal evolution. Unfortunately, there is currently no direct way to constrain the speed of early lunar tidal evolution. While the age of the Earth-Moon system implies Earth's long-term average tidal $Q_E \simeq 34$, the current tidal $Q$ of Earth is about 12, meaning that there must have been at least some intervals of higher tidal $Q$ in the past. \citet{bil99} show that often contradictory claims of tidal periods captured in the geological record only concern the last billion years or so, and the timeline of the early history of the Earth-Moon system is completely unconstrained. Some models of ocean dissipation \citep{web82} indicate general reduction of $Q$ over geological time, with tidal $Q$ being a factor of several higher than now in the first billion years of the Solar System, but so far there is no widely accepted correspondence between lunar distance and time.

 \begin{table}
 \caption{Approximate times when specific lunar semimajor axes were reached for two different values of early Earth's tidal quality factor: $Q=34$ (4.5 Gyr average) and $Q=100$. For simplicity, we assumed that Earth had a Love number $k_2^E=0.3$ (like today), and that the Moon formed exactly 4.5 Gyr ago.}
 \label{table1}	
 \centering
 \begin{tabular}{l c c}
 \hline
  Lunar $a$& Time for $Q_E=34$ & Time for $Q_E=100$  \\
 \hline
   30$R_E$  & 4.45 Gya & 4.35 Gya\\	
   35$R_E$  & 4.4 Gya & 4.1 Gya\\
   40$R_E$  & 4.2 Gya & 3.6 Gya\\
   45$R_E$  & 3.8 Gya & 2.5 Gya\\
 \hline
\end{tabular}
\end{table}

In Table \ref{table1}, we present two possible timelines for lunar tidal evolution as a function of Earth's tidal $Q$ during the system's early history. These numbers do not establish a preferred time-obliquity correspondence, but rather illustrate that a wide range of reasonable parameters would predict a large lunar core-mantle obliquity at the epoch when the lunar magnetic field is thought to have been strongest, 4.25-3.6 Gyr ago \citep{tik14, wei14}. In Table \ref{table1} we used $\dot{a} \simeq a^{-11/2} Q_E^{-1}$ \citep{md99}, i.e. we assumed a constant $Q$ for early Earth, and ignored lunar eccentricity and obliquity tides, as well as temporary halts of tidal evolution due to the Laplace Plane transition \citep{cuk16}. 

The obliquities shown in Figs. \ref{cassini}-\ref{core2} are model-dependent, and lunar tidal evolution models that produced lower past lunar inclinations \citep{tou94b} would also result in smaller obliquities (although high obliquities close to the Cassini State transition are inevitable). However, lower past lunar inclinations in models predating work by \citet{che13} are primarily an artifact of the obliquity tides being ignored. While \citet{pea78} considered tidal heating from obliquity tides, they did not investigate associated inclination damping; \citet{wil01} proposed that tidal heating during the Cassini State transition powered the lunar dynamo, but did not address orbital consequences. Therefore, pre-2013 models of lunar tidal evolution cannot be directly used to derive past lunar mantle obliquities. More recently, the model of \citet{pah15} gets around the inclination damping at the Cassini State transition by exciting lunar inclination through planetesimal flybys after a rapid early lunar orbital expansion. Since \citet{pah15} propose that the Moon underwent a rapid orbital expansion in the first 10 Myr, and had a close-to-present inclination at $a \simeq 45 R_E$ after the encounters were over ($\simeq 100$~Myr), it appears unlikely the Moon would have had large enough obliquity to power a dynamo at the ``magnetic epoch'' 4.25-3.6 Gya. While the effects of late planetesimal encounters need more exploration (we ignored them in this work as they happen before the Cassini State transition for tidal parameters we considered), it is likely that the \citet{pah15} model is incompatible with a precession-driven dynamo, and that a different mechanism would be needed to power the ancient lunar dynamo.

In our calculations, we ignored the presence of the solid inner core. While the size of the inner core is poorly constrained \citep{wil14}, it should have significant effects on the core dynamics by introducing new nutation frequencies \citep{dum16} and it may in some cases significantly change the tilt of the outer core \citep{sty18}. It is thought that the solid inner core grew over time \citep{sch15}, so it would have been much smaller (or not present at all) during the early history of the Earth-Moon system that we study here. However, assuming a later formation for the inner core may not be viable, as some models of the magnetic dynamo generation require the presence of the solid inner core \citep{sta17}, so improved dynamical models will have to include a two-component core. Based on our results for the outer core, the inner core is unlikely to be synchronously locked as \citet{sta17} have proposed, as friction with the sub-synchronous outer core would likely be a more important factor for determining the inner core's spin than Earth's tidal torque. Additionally, \citet{sty18} have found that the free inner core nutation period is currently in the  10-40~yr range, making a large forced tilt due to resonant interaction with the 18.6~yr mantle precession period likely.

Finally, we note that a magnetic field that is locked to the core, or changes its geometry relatively slowly relative to the core, would not be stationary relative to the lunar mantle. Earth's core is practically co-rotating with the mantle and its spin axis is not significantly offset from that of the mantle (as the free core nutation is much, much faster than Earth's axial precession). Therefore, experience from Earth is not directly applicable to the Moon, and a lunar dipole field generated in the core would drift across the lunar surface due to different rotations of the core and the mantle. For example, at $a=40 R_E$, when we find the core-mantle obliquity to be 25$^{\circ}$, a magnetic pole that is aligned with the core spin axis would sweep on the surface a circle at $65^{\circ}$ latitude in a course of a lunar orbital period ($\simeq15$ days at that time). The pole of a dipole field that is significantly tilted relative to the core spin axis would trace a much more complex track on the timescales determined by the differential rotation of the core and the mantle. Therefore, directional magnetization of lunar rocks does not record a long-term orientation of the lunar magnetic field, but possibly only a snapshot taken at the critical points of the material's thermal evolution. Some of the recent work on the locations of proposed paleopoles does indicate a great variability in the inferred orientation of the lunar dynamo \citep{oli17, nay17}, and relative motions of the core and the mantle may at least partially explain these findings.  

\section{Conclusions}

In this work we numerically modeled the rotational dynamics of the Moon's core and the mantle during the early history of the Earth-Moon system, and we report the following conclusions:

1. The lunar core was precessing independently of the mantle for practically all of lunar history.

2. Our numerical model also suggests that in the distant past the lunar core rotated significantly slower than the mantle, assuming that the main dissipative torque on the core comes from friction at the core-mantle boundary.

3. The presence of the small lunar core makes the non-synchronous rotation before and after the Cassini State transition even more likely, and the CMB friction should have caused additional damping of lunar inclination, separate from obliquity tides.

4. For much of the early history of the Earth-Moon system there was a large mutual obliquity between the core and the mantle. This has direct implications for the generation of the ancient lunar magnetic field, and we discuss plausible absolute timings of events in lunar tidal evolution.

Finally, we note that this is only the first step in numerically modeling the dynamics of the lunar core. Due to numerical limitations, we only explored stationary solutions for the lunar rotation, and in the future we hope to produce a self-consistent lunar tidal evolution model that includes the effects of core-mantle interactions. We also hope that additional factors such as the presence of the solid inner core and the possible non-rigid behavior of the core-mantle boundary will be addressed in future work. 

\acknowledgments
M\' C is supported by NASA Emerging Worlds award NNX15AH65G. We thank Mathieu Dumberry and an anonymous reviewer for their extremely helpful reviews of the previous version of this paper. The source code for our integrator {\sc CR-SISTEM} is available through the Astrophysics Source Code Library.

%% ------------------------------------------------------------------------ %%
%% References and Citations

%%%%%%%%%%%%%%%%%%%%%%%%%%%%%%%%%%%%%%%%%%%%%%%
% BibTeX is preferred:
%
\bibliography{refs.bib}
%
% don't specify 
%%%%%%%%%%%%%%%%%%%%%%%%%%%%%%%%%%%%%%%%%%%%%%%

\end{document}